\documentclass[preprint]{article}
\usepackage{latexsym}
\usepackage{graphicx}
\begin{document}
\title{
Evolution of genetic code through isologous diversification of cellular states
}
\author{
Hiroaki Takagi$^{1}$, Kunihiko Kaneko$^{1}$
\and Tetsuya Yomo$^{2,1}$\\
$^{1}${\small \sl Department of Pure and Applied Sciences}\\
{\small \sl University of Tokyo, Komaba, Meguro-ku, Tokyo 153, JAPAN}\\
$^{2}${\small \sl Department of Biotechnology}\\
{\small \sl Faculty of Engineering}\\
{\small \sl Osaka University 2-1 Suita, Osaka 565, JAPAN}
}

\maketitle

\begin{abstract}      
Evolution of genetic code is studied as the change in the
choice of enzymes that are used to synthesize amino acids
from the genetic information of nucleic acids. We propose the following theory: the differentiation of physiological states of a cell allows for the different choice of enzymes, and this choice is later fixed genetically through evolution.  
To demonstrate this theory, a dynamical systems model consisting of the concentrations of 
metabolites, enzymes, amino acyl tRNA synthetase,
and tRNA-amino acid complex in a cell is introduced and numerically studied.
It is shown that the biochemical states of cells are differentiated by cell-cell interaction, and each differentiated type takes to use different synthetase.
Through the mutation of genes, this difference in the genetic code is amplified
and stabilized.  Relevance of this theory to the evolution of non-universal genetic
code in mitochondria is suggested.

The present theory for the evolution of genetic code is based on
our recent theory of isologous symbiotic speciation, which is briefly reviewed.
According to the theory, phenotypes of organisms 
are first differentiated into distinct types through the interaction and developmental dynamics, 
even though they have identical genotypes, and later
with the mutation in genotype,
the genotype also differentiates into discrete types, while
maintaining the `symbiotic' relationship between the types.
Relevance of the theory to natural  
as well as artificial evolution is discussed.

\end{abstract}

\section{Introduction}

The protein synthetic system adopted in today's living organisms has
a very large and complex network. It consists of over 120 kinds of molecules, such as tRNA, ARS(aminoacyl tRNA synthetase), mRNA, 20 kinds of amino acids, ribosome, ATP, etc. In this system, genetic code plays an important role to link genetic information in DNA to phenotypic functions, 
where genetic code was considered to be stable.  From such considerations and experimental results, the genetic code was once considered to be universal,  and ``frozen accident theory"
was proposed  by F.Crick\cite{F.Crick}, in which the 
genetic code is assumed to be fixed by frozen accident in 
the early history of life. From recent studies, however, several non-universal genetic codes
 were found, for example,  in mitochondrial DNA.
Now it is recognized that genetic code is not universal and can change in a long term. 
Considering these stability and flexibility of the genetic code, 
it is important to study the evolution of genetic codes with these two aspects,
which might look like contradicting superficially. 

To discuss the evolution of genetic codes, it is necessary to point out two basic features 
of genetic  codes. 

The first point concerns about the relationship between genetic codes and the molecular structure. Although tight chemical coupling between codon and amino acid such as a key-keyhole relationship was initially assumed, it is now believed
that there is no specific interaction between codon and amino acid\cite{JMS}. Ueda et al. have recently discovered ``polysemous  codon'' in certain Candida species,
where two distinct amino acids are assigned by a single codon\cite{T.Ueda}.  Now
 looseness in genetic code is seriously studied.

Second, the evolutionary change of genetic codes has also been studied after the discovery of 
non-universal genetic code. Among these studies, ``codon capture theory'', proposed by 
Osawa $\&$ Jukes is most popular\cite{S.Osawa}. The essence of the theory is as follows:
If some change to genetic code occurred without any intermediate stage,  a sense codon 
would be changed to a nonsense one, which would cause vital damage to the survival.
Therefore, it is necessary to pass through some intermediate stage in evolution, during which
the change of genetic code is not fatal.  If genetic code is degenerate and some specific 
triplet is hardly used, tRNA and ARS that correspond to the specific triplet 
can change their coding without fatal damage. 

With these two points in mind, we consider the problem of evolution of genetic codes.
First, we expect that genetic code must have passed through the stage with
some ambiguity or looseness in the course of the evolution,
since otherwise it is hard to imagine that the genetic code has evolved 
without having a fatal damage to an organism.  Then, how is such looseness supported?
How is a different coding for the translation supported biochemically?
If the difference in genetic codes were solely determined by a
genetic system all through the evolutionary process, it would be difficult to consider
how the change from one code to another could occur smoothly, without a fatal 
damage to a cell.  Instead,
we propose here that the difference in the translation is not solely determined
by the nucleus, but is also influenced by the physiological state of a cell,
at least at some stage of evolution.   Indeed, as will be shown, it is rather plausible that cells with identical genes
can have different physiological states.  Such differentiation
is expected to occur according to the ``isologous diversification theory", proposed 
for cell differentiation\cite{KK94,KY97,KY99,FK98,FK98b,KF00}. 
Since the translation system from nucleic acid to amino acid is influenced by several enzymes within a cell,
the difference in the physiological state can introduce some change in
the translation  also.  By constructing a model with
several biochemicals, we will give an example with non-unique
correspondences from nucleic acids  to amino acids.

The present paper is organized as follows.  In \S 2, we describe
isologous speciation theory in some detail, since it gives
a basis for the present theory of evolution of genetic codes.
In \S 3, we introduce our model of a cell with several chemicals.
by choosing such biochemical reactions that allow for differentiation in physiological states and ARS.  In \S 4, we take into account the mutation 
into a genetic system, and study
how different genetic codes are established through the evolution.
Through the extensive simulation, we propose the following
theory for the evolution of genetic code:
first, phenotypic differentiation occurs for metabolic dynamics through cell-cell
interaction. Then each differentiated group of cells starts to use different ARS,
and adopt a different way in translating nucleic acid to protein (enzyme). 
Then, through evolutionary process with competition for reproduction and mutation to genes, 
this difference in physiological state results in a difference in genes, and
one-to-one correspondence is established between differentiated phenotype
and mutated genes, so that each group can clearly be separated both in phenotype and genotype.
After this evolutionary process, the difference in the translation is
fixed.  Each group finally achieves a different genetic code, that is now fixed in time,
and the initial ambiguity or looseness in coding is reduced.
Summary and discussion on the relevance of the present result
to cell biology as well as
to artificial life are given in \S 5.

\section{Isologous Speciation Theory}

The background for the present theory for the evolution of genetic
codes lies in our isologous symbiotic sympatric speciation theory\cite{KY00,KY-AL}.
Since the theory is essential to the present study, we
explain it at length in the present section\footnote{
This section is somewhat independent of other parts,
and one can skip it or read only this section.}.

\subsection{Background of the isologous symbiotic sympatric speciation}

The question why organisms are separated into distinct
groups, rather than exhibiting a continuous range of characteristics,
originally raised by Darwin\cite{Darwin}, has not yet been fully answered,
in spite of several attempts to explain sympatric speciation.
Difficulty in stable sympatric speciation,
i.e., process to form distinct groups with reproductive isolation,
lies in the lack of a known
clear mechanism how two groups, which have just started to be separated, 
coexist in the presence of mutual interaction and mixing of genes by mating.  
So far people try to propose
some mechanism so that the two groups do not mix and survive independently,
as is seen in sexual isolation by mating preference 
(e.g., \cite{JMS2,Lande,Turner,book,Kond,Dieck}).
However, this type of theory cannot answer how such mating preference that is
`convenient' for sympatric speciation, is selected.
Furthermore, if one group may disappear by fluctuations
due to finite-size population, the other group does not reappear.
Coexistence of one group is not necessary for the survival of the other.
Hence the speciation process is rather weak against possible fluctuations
that should exist in a population of finite size.

Of course, if the two groups were in a symbiotic state, 
the coexistence would be necessary for the
survival of each.  However, as long as the phenotype is
a single-valued function of genotype, two groups 
with little difference in genotypes must have almost same phenotypes. 
Hence, in the beginning of speciation process,
it might be hard to imagine such a `symbiotic' mechanism.  Accordingly, it is
generally believed that sympatric speciation, stable against fluctuations, is
rather difficult. 

Recall the standard standpoint for the evolution in 
the present biology\cite{Futsuyma,Alberts}. (i) First,
each organism has genotype and phenotype.
(ii) Then, the fitness for survival is given for a phenotype, and
Darwinian selection process acts for the survival
of organisms, to have a higher fitness (iii) Only the genotype is transferred to 
the next generation (Weissman's doctrine) (iv) Finally, there is a 
direct flow only from a genotype to phenotype, i.e.,  a phenotype is determined through 
developmental process,  given a genotype and environment ( the central dogma of
molecular biology).  Although there may be some doubt
in (iii) (and (iv)) for some cases, we follow this standard
viewpoint here.  

Note, however, that (iv) does not necessarily mean that the phenotype is
`uniquely determined'.  In the standard population genetics, this uniqueness is 
assumed, but it is not necessarily postulated within the above
standard framework.
Indeed, an answer for the speciation problem is provided
by dropping this assumption and taking the isologous diversification.
Furthermore, there are three reasons to make us doubt this assumption of
the uniqueness.

First, we have previously proposed isologous diversification
theory, where two groups with distinct phenotypes appear even from 
the same genotype\cite{KK94,KY97,KY99,FK98,FK98b,KF00}.
In this theory, due to the orbital instability in developmental process,
any small difference (or fluctuation) is amplified to a macroscopic level,
so that the dynamical state of two organisms (cells) can be different,
even if they have a same set of genes.  The organisms are differentiated into
discrete types through the interaction, where
existence of each type is necessary to eliminate the dynamic instability in 
developmental process,  which underlies when the ensemble of
one of the types is isolated.
Here, existence of each type is required for the survival of each other,
even though every individual has identical, or slightly different genotypes.

Second, it is well known experimentally that in
some mutants, various phenotypes arise from a single
genotype, with some probability\cite{Holmes}. This phenomenon is known as 
low or incomplete penetrance\cite{Opitz}. 

Last, the interaction-induced phenotypic diversification is 
clearly demonstrated in an experiment, 
for specific mutants of {\it E. coli}.  In fact, the coexistence of
(at least) two distinct types of enzyme activity is demonstrated, 
in a well stirred environment of a chemostat,
although they have identical genes \cite{Yomo,Kashi}.
Here, when one type of {\it E. coli} is
removed externally, the remained type starts to differentiate again
to recover the coexistence of the original two types.
It is now demonstrated that distinct
phenotypes (as for enzyme activity) appear,
according to the interaction among the organisms, even though they
have identical genes. 

Hence, we take this interaction-induced phenotypic 
differentiation from a single genotype seriously into account and discuss its relevance to evolution.
For it, we have to consider a developmental process that maps a
genotype to a phenotype.  Consider 
for example an organism with several biochemical processes.
Each organism possesses such internal dynamic processes
which transfer external resources into some products depending on the internal
dynamics.  

Here, the phenotype is represented by a set of
variables, corresponding to biochemical processes.
Genes, since they are nothing but information expressed on DNA, could in
principle be included in the set of variables. However, according to the
central dogma of molecular biology (requisite (iv)), the gene has a 
special role among such variables. Genes can affect phenotypes, 
the set of variables, but the
phenotypes cannot change the code of genes. During the life cycle, changes
in genes are negligible compared with those of the phenotypic variables
they control.  
In terms of dynamical systems, the genes
can be represented by control parameters that govern the dynamics of phenotypes, since
the parameters in an equation are not changed through the developmental process, 
while the parameters control the dynamics of phenotypic variables.
Accordingly, we represent the genotype by a set of parameters. 
When an individual organism is reproduced, this set of
parameters changes slightly by mutation.

Next, there are interactions between individuals through exchange of
chemicals.  Some chemicals secreted out from one organism may be taken by
another, while they have competitive interactions for resources. This interaction depends on the internal state of the unit. In dynamical
systems theory, the interaction term is introduced for the change
of (biochemical) states of the unit.

Then, each individual replicates when some chemicals are accumulated
after chemical reactions. 
Since, genotypes
are given by a set of parameters representing the 
biochemical reaction, they slightly mutate by reproduction.
With each replication, the parameters are changed slightly 
by adding a small random number.

As the number of organisms grow, not
all of them generally survive.
This competition for survival is included by random removal of organisms 
at some rate as well as by a given depending on their (biochemical) state. 

\subsection{Theory for the speciation}

We have carried out simulations of several models
of the above type, from which
a speciation theory is proposed, as
is described as follows \cite{KY00,KY-AL}.
(see Fig.1 for schematic representation).

{\bf Stage-1: Interaction-induced phenotypic differentiation}

When many individuals interact competing for finite resources, the
phenotypic dynamics start to be differentiated even though the genotypes
are identical or differ only slightly. This differentiation generally
appears if nonlinearity is involved in the internal dynamics of some
phenotypic variables. Slight differences in variables between
individuals are amplified by the internal dynamics (e.g., metabolic
reaction dynamics). Through interaction between organisms, the difference
in phenotypic dynamics are amplified and the phenotype states tend to be 
grouped into two (or more) types.
The dynamical systems mechanism for such differentiation
was first discussed as clustering \cite{KK90}, and then extended,
to study the cell differentiation
\cite{KK94,KY97,KY99,FK98,FK98b,KF00}. In fact, the orbits 
lie in a distinct region in the
phase space, depending on each of the two groups that the individual $i$ belongs to. 
Note that the difference at this stage is fixed neither in the genotype nor in  the phenotype. The progeny of a reproducing individual may
belong to a distinct type from the parent. 
If a group of one type is removed, then some individuals of the other type 
change their type to compensate for the missing type.
To discuss the present mechanism in biological terms, consider a given 
group of organisms
faced with a new environment and not yet specialized for the processing of
certain specific resources.
Each organism has metabolic (or other ) processes with a biochemical
network.  As the number of organisms increases, they compete for
resources.  As this competition  becomes stronger, the 
phenotypes become diversified to allow for different uses in metabolic cycles, 
and they split into two (or several) groups.  Each group is specialized in 
processing of some resources.  
Here, the two groups realize differentiation of
roles and form a symbiotic relationship. Each group is
regarded as specialized in a different niche, which is provided by another group.

{\bf Stage-2: Co-evolution of the two groups to amplify the difference of.
genotypes}

At the second stage of our speciation, difference in
both genotypes and phenotypes is amplified. This is realized by a kind of positive
feedback process between the changes in geno- and phenotypes.
In general, there is a parameter which has opposite influence on the growth 
speed between the two phenotypes.  For example, 
for the upper group in Fig. 1b), assume that the growth speed is higher when
the parameter is larger, and the other way around for the lower group.
Then, through the mutation and selection, genetic parameters of the two
phenotype groups start to separate as shown in Fig.1c).

Indeed, such parameter dependence is not exceptional. As a simple
illustration, the use of metabolic processes is different
between the two groups. If the upper group uses one metabolic cycle more,
then the mutational change of a specific parameter to enhance the use of
the cycle is in favor for the upper group, while the change to reduce it
may be in favor for the lower group. Indeed, several numerical results
support that there always exist such parameters.
This dependence of growth on genotypes leads to genetic separation of the
two groups.

With this separation of two groups, each phenotype (and genotype) tends to
be preserved by the offspring, in contrast with the first stage. Now, 
distinct groups with recursive reproduction
have been formed. However, up to this stage, 
the two groups with different phenotypes cannot exist by
themselves in isolation. When isolated, offspring with the phenotype of the
other group start to appear. 
The developmental dynamics in each group, when isolated,
are unstable and some individuals start to be differentiated to recover
the other group.  The dynamics, accordingly each phenotype,
is stabilized by each other through the interaction.  Hence, two groups
are in a symbiotic state.
To have such stabilization,
the population of each group has to be balanced.
Even under random fluctuation
by finite-size populations and mutation, the population balance of each group
is not destroyed. 
Accordingly, our mechanism of genetic diversification is robust
against perturbations.

{\bf Stage-3 Genetic fixation and isolation of differentiated groups}

Complete fixation of the diversification to genes occurs at this
stage.  Here, even if one group of units is isolated, the offspring
of the phenotype of the other group are no longer produced. Offspring of
each group keep their phenotype (and genotype) on their own. This is
confirmed by numerically eliminating one group of units.

Now, each group has one phenotype corresponding to each genotype, even without
interaction with the other group. Hence, each group is a
distinct independent reproductive unit at this stage. This stabilization 
of a phenotypic state is 
possible since the developmental
flexibility at the first stage is lost, due to the shift
of genotype parameters. The initial phenotypic change introduced by the
interaction is now fixed to genes. 

To check the third stage of our theory, it is straightforward to study
the further evolutionary process from only one isolated group. In order to do this,
we pick out some population of units only of one type,
after the genetic fixation is completed and both the geno- and phenotypes are separated into two groups, and start the simulation again.  
When the groups are picked at this third stage,
the offspring keep the same phenotype and genotype. Now, only one
of the two groups exists. Here, the other group is no longer necessary to
maintain stability. 

\subsection{Some remarks}

To check the condition for speciation, we have performed numerical 
experiments of evolution, 
by choosing model parameters so that differentiation into two distinct 
phenotypic groups
does not occur initially.  In this case, separation into 
two (or more) groups with distinct pheno/geno-types is never observed,
even if the initial variance of
genotypes is large, or even if a large mutation rate is adopted.

Next, the genetic differentiation always occurs 
when the phenotype
differentiates into two (or more) distinct groups,
as long as mutation exists. Hence, phenotypic
differentiation is a necessary and sufficient
condition for the speciation under a
standard biological situation, i.e., a process with reproduction, mutation, and
a proper genotype-phenotype relationship. 
Note that the interaction-induced phenotypic differentiation is
deterministic in nature.
Once the initial parameters of the
model are chosen, it is already determined whether such differentiation 
will occur or not. 

The speciation process is also stable against sexual 
recombination.  In sexual recombinations, two genes are mixed, and the
differentiated two groups may be mixed and the speciation may be 
destroyed.  We have found that our speciation process is stable
under sexual recombinations\cite{KY00,KY-AL}.
Indeed, the hybrid are formed with random mating, but they 
have lower reproduction rate, and finally they become sterile.
Thus the definition of species, i.e., sterility of the hybrid, is resulted.

In our speciation process, the potentiality for a single genotype to
produce several phenotypes decreases. After the
phenotypic diversification of a single genotype, each genotype again
appears through mutation and assumes one of the diversified phenotypes in the
population. Thus the one-to-many correspondence between the original
genotype and phenotypes eventually ceases to exist. As a result, one may expect
that a phenotype is uniquely determined for a single genotype in
wild types, since most
organisms at the present time have gone through several speciation
processes.  

Finally, it should again be stressed that {\sl neither any Lamarckian mechanism
nor epigenetic inheritance is assumed} in our theory, in spite of the
genetic fixation of the phenotypic differentiation. 
Only the standard flow from genotype to
phenotype is included in our theory.  Note also that 
genetic `takeover' of phenotype change was also proposed
by Waddington as genetic assimilation\cite{Waddington}, in possible 
relationship with Baldwin's effect.  Using the idea of epigenetic landscape,
he showed that genetic fixation of the displacement of phenotypic character
is fixed to genes.  In our case the phenotype differentiation is
not given by `epigenetic landscape',
but rather, the developmental process forms different characters through the
interaction.  Distinct characters are stabilized through the interaction.
With this interaction dependence, the two groups are necessary with each other,
and robust speciation process is possible.   

\section{Our Model for the Evolution of Genetic Code}

Now, let us come back to the problem of the evolution of genetic code.
Here, we construct an abstract model to demonstrate the theory for the
evolution of genetic code.
Of course, it is almost impossible to describe all factors of complex cellular process.  Furthermore, even if we succeeded in it, we 
could not understand how the model works, since the model is too much complicated.
Rather, we extract only some basic features of a problem in concern,
and construct a model to understand a general aspect of the evolution of the 
genetic code.   
In particular, we show how differentiated ``phenotypes" are organized,
that adopt a different coding in the translation from nucleic acids to 
amino acids, based on the isologous diversification.
Then, with the evolution with
mutation of genes, the different translations will be shown to be
established following the theory of the last section.  

We start from a cell with a set of variety of biochemicals.  Considering 
the metabolic and genetic process, at least four kinds of
basic compounds are necessary, namely, metabolic chemicals, enzymes for metabolic reaction, 
chemicals for genetic information,
and enzymes to make translation of genetic information to protein.  
In the present paper, these four kinds of chemicals are chosen as 
the metabolites (metabolic chemicals), enzymes for metabolites,  tRNA-amino acid complexes, and ARS,
respectively for this set of chemicals.  Now,
as a state variable characterizing the cell, we introduce\\
\\
$c_i^{(j)}(t)$: concentration of $j$th metabolic chemical in $i$th cell  \\
$a_i^{(j)}(t)$: concentration of $j$th enzyme for metabolites in $i$th cell  \\
$e_i^{(j)}(t)$: concentration of $j$th ARS in i's cell  \\
$x_i^{(j)}(t)$: concentration of $j$th tRNA-amino acid complexes in $i$th cell  \\
\\
As for the dynamics of these chemicals, we consider the following processes.

\begin{itemize}
\item intra-cellular chemical reaction network
\item inter-cellular interaction
\item cell division and mutation
\end{itemize}

Now, we describe each process. See Fig.2 and Fig.3 for schematic representation of our model.

\subsection{Intra-cellular chemical reaction network}

In general, each biochemical reaction in cells is catalyzed by some enzymes.
Here, each metabolic reaction is assumed to be catalyzed by each specific enzyme, and a simple form of reaction rate is adopted given by just the product of the concentrations of the substrate and enzyme in concern.  (This
specific form is not essential, and the same qualitative results are obtained
 by using some other form, such as Michaelis-Menten's one.)   
Here we choose a network consisting of reactions
from some metabolite $j$ to other metabolite $k$ catalyzed by the enzyme $k$.  The network is
chosen randomly, and is represented by  a reaction matrix $W(j,k)$, which takes
1 if there is a reaction path, and 0 otherwise.  The network is fixed throughout the simulation.
Of course, the dynamics can depend on the choice of the reaction network.  Here we choose such network that allows for some oscillatory dynamics. The oscillation
is rather commonly observed, as long as there
is a sufficient number of autocatalytic paths.

Next, each enzyme, including ARS, that for the synthesis for tRNA, is synthesized from amino acids.
This synthesis is again catalyzed  by some enzyme.  
This synthesis is given by a resource table.
For the enzyme $a^{(j)}$ and the ARS $e^{(j)}$, the tables  are given by
 $V(j,k)$ and  $U(j,k)$ respectively. We also set all entries of $V(j,k)$, $U(j,k)$ at random, by keeping a normalization
${\sum_k}V(j,k)={\sum_k}U(j,k)=1.0$.

Third, we assume that ARS produces tRNA-amino acid complexes in proportion to its amount.
The correspondence between the two is given by
a reaction matrix $T(j,k)$, which is 1 if ARS $e^{(j)}$ produces tRNA-amino acid $x^{(k)}$, 
and 0 otherwise. To include ambiguity,  we allow one to many correspondence
between $e$ to $x$. The matrix $T(j,k)$ is again chosen  randomly.

Here we take $P(=16)$ species of metabolic chemicals (C) and the 
corresponding enzymes (A),
$R(=12)$ species of ARS (E), and $Q(=6)$ tRNA-amino acids(X). 
Accordingly the concentration change of chemicals by intracellular process
is given by

\begin{eqnarray}
dc_i^{(j)}(t)/dt= D_1\sum_{k=1}^{P}{W_i^{(k,j)} a_i^{(k)}(t)c_i^{(k)}(t)}\nonumber\\
               -D_1\sum_{k=1}^{P}{W_i^{(j,k)} a_i^{(j)}(t)c_i^{(j)}(t)}\nonumber
\end{eqnarray}
\begin{eqnarray}
da_i^{(j)}(t)/dt = D_3(\sum_{k=1}^{Q} V_i^{(j,k)} x_i^{(k)}(t)) a_i^{(j)}(t) c_i^{(l(j))}(t)\nonumber
\end{eqnarray}
\begin{center}
(where $l(j)$:$j$$\rightarrow$$l$ gives a one-to-one mapping.)
\end{center}
\begin{eqnarray}
de_i^{(j)}(t)/dt = D_4(\sum_{k=1}^{Q} U_i^{(j,k)} x_i^{(k)}(t)) a_i^{(m(j))}(t) c_i^{(n(j))}(t)\nonumber
\end{eqnarray}
\begin{center}
(where $m(j)$,$n(j)$:$j$$\rightarrow$$l$ give one-to-one mappings.)
\end{center}

Finally, tRNA-amino acid complexes, which provide the materials of all enzymes,
are assumed to change with a faster time scale than the above three types of chemicals.
Hence, we adiabatically eliminate its concentration to give the equation for it.
By setting 

\begin{eqnarray}
dx_i^{(j)}(t)/dt=D_5\sum_{k=1}^{R} T_i^{(k,j)} e_i^{(k)}(t)\nonumber\\
 -D_3 x_i^{(j)}(t)\sum_{k=1}^{P} V_i^{(k,j)} a_i^{(k)}(t) c_i^{(l(k))}(t)\nonumber\\
 - D_4 x_i^{(j)}(t)\sum_{k=1}^{R} U_i^{(k,j)} a_i^{(m(k))}(t) c_i^{(n(k))}(t)=0,\nonumber
\end{eqnarray}
we obtain
\begin{eqnarray}
x_i^{(j)}(t)=D_5\sum_{k=1}^{R} T_i^{(k,j)} e_i^{(k)}(t)\nonumber\\
/\{D_3\sum_{k=1}^{P} V_i^{(k,j)} a_i^{(k)}(t) c_i^{(l(k))}(t) \nonumber\\
+ D_4\sum_{k=1}^{R} U_i^{(k,j)} a_i^{(m(k))}(t) c_i^{(n(k))}(t)\}\nonumber
\end{eqnarray}
 
Note that 
the translation process of genetic information to proteins is given by the
process between $X$ (tRNA-amino acid complexes) and $E$ (ARS).   One can discuss the difference in coding
by examining which species of $E$ has nonzero concentration, and acts in the
translation process.  We first study how the difference in physiological states
given by $C$ affects in the choice of $E$, for our purpose of the problem.

\subsection{Cell-cell interaction}

According to the isologous diversification theory, cell-cell interaction is
essential to establish distinct cell states.  Here we consider the interaction
as diffusion of some chemicals through the medium.
In this model, we assume that only metabolic chemicals ($c$) are
transported through the membrane, which is rather plausible biologically.
Assuming that cells are in a completely stirred medium,
we neglect spatial variation of chemical concentrations in the
medium. Hence we need only another set of concentration variables\\
\\
$C^{(j)}(t)$: concentration of $j$th metabolic chemical in the medium \\
\\
Therefore, all the cells interact with each other through the same environment.
As a transport process we choose the simplest diffusion process, i.e.,
a flow proportional to the concentration difference between the inside and outside of a cell. 

Of course, the diffusion coefficient depends on the metabolic chemical.
Here,  for simplicity we assume that all the chemicals $c$ are classified
into either penetrable or impenetrable ones.  The former has
the same diffusion coefficient $D_2$, while for the latter the coefficient is
set to 0. Here we define the index ${\sigma_m}$, 
which takes 1 if a chemical $c^{(m)}$ can penetrate the membrane, and otherwise 0. 
Each cell grows by taking in penetrable chemicals from the medium
and transforms them to other impenetrable chemicals. \\

Accordingly, the term for the diffusion 
\begin{eqnarray}
{\sigma_j}D_2(C^{(j)}(t)-c_i^{(j)}(t))
\end{eqnarray}
is added to the equation for $dc_i^{(j)}(t)/dt$,
while the concentration change in the medium is given by
\begin{eqnarray}
dC^{(j)}(t)/dt=& &\nonumber\\ 
D_6(\overline{C^{(j)}}-C^{(j)}(t))&-&{\sigma_j}D_2\frac{(\sum_{k=1}^{N} C^{(j)}(t)-c_k^{(j)}(t))}{Vol},\nonumber
\end{eqnarray}
where the parameter $Vol$ is the volume ratio of a medium to a cell, and $N$ is the number of cells.  Since
these chemicals in the medium are consumed by a cell, we impose a flow of penetrable chemicals from 
the outside of the medium, that is proportional to the concentration differences.
This term is given by the term $D_6(\overline{C^{(j)}}-C^{(j)}(t))$,
where the external concentration of chemicals $C^{(j)}$ is denoted by 
$\overline{C^{(j)}}$.

The variables $c_i^{(j)}$, $a_i^{(j)}$, $e_i^{(j)}$, and $x_i^{(j)}$ stand for the concentrations.  Since the volume
of a  cell can change with a flow of metabolites, its change should be taken into
account.  Here, we compute the increase of the volume from the flow of chemicals
by the sum of the term in eq.(1).  The concentration is diluted in accordance with
this increase of the volume.  With this process the sum
\begin{eqnarray}
   \sum_{j=1}^{P} c_i^{(j)}(t) + \sum_{j=1}^{P} a_i^{(j)}(t) + \sum_{j=1}^{R} e_i^{(j)}(t) \nonumber
\end{eqnarray}
is preserved through the development of a cell, and the sum is set at 1 here.

\subsection{Cell division}

Each cell gets resources from the medium and grows by changing them to other chemicals.  With the flow into a cell, the chemicals are accumulated
in each cell.  As mentioned, this leads to the increase of the volume of
a cell.  We assume that
the cell divides, when the volume is twice the original.
After the division, the volume of each cell is set to be half.
In the division process, a cell is divided into
two almost equal cells, with some fluctuations.
Hence, the concentrations of chemicals
$b_i^{(j)}$ (where $b$ represents either  $c$, $a$, or $e$) are
divided into $(1+\eta)b_i^{(j)}$ and $(1-\eta)b_i^{(j)}$,
with $\eta$ as a random number over $[-10^{-2},10^{-2}]$.
As will be shown later, this
fluctuation can be amplified to a macroscopic level.
The amplitude is not essential, but the existence of fluctuation itself
is relevant to have differentiation.

\subsection{Mutation}

 To discuss the evolution of genetic codes in a long run, we need
 to include mutation to genes.  In our model, the genetic information
 is translated from DNA into amino acid.  Here both $U$ and $V$ are changed 
by the mutation to the table of enzyme.  At each division,
each element of the matrix  $U$ or $V$ is mutated
by a random number $\kappa$ with the range of
$[-\varepsilon , \varepsilon]$, where $\varepsilon$ corresponds to
the amplitude of the mutation rate, which  we later set at
 $10^{-3}$ for most simulations. 

Note that this matrix corresponds to genotype, while
other chemical concentrations give biochemical states of the cell.
Since in our model, there is no direct process to change the matrix from
the concentrations,  the ``central dogma" of the molecular biology
is satisfied, i.e., genotype can change phenotype, but not otherwise.
We also assume that mutation to genes affects only to the catalytic 
abilities of enzymes $a$, $e$, and not to the specificity of catalytic
reactions. \\

Recall that the difference in the genetic code
is represented by which kinds of ARS are used in a cell,
depending on the  physiological state of the cell.
Here, we are interested in how this difference is fixed genetically
through the evolution.  
With the change of the matrix element of $U$ corresponding to the ARS,
the use of specific kind of ARS may start to be fixed,
with the increase of some matrix element (to approach unity),
according to the theory of \S 2. If this is the case,
specific mappings between ARS and tRNA-amino acid 
complex are selected, to establish a different coding system.
We will confirm this argument in the
following sections, based on the simulation results of our model.

\section{Isologous Diversification of the Genetic Codes}

First, we discuss the behavior of the present cell system, without 
introducing mutation.
We assume that intracellular chemical dynamics for a single cell system,
show oscillation. Since there are many oscillatory reactions in real cells 
such as Ca$^{2+}$, cAMP, NADH, and the oscillation is easily brought about
 by autocatalytic reactions (as also given by the hypercycle \cite{M.Eigen&P.Shuster}), the existence of oscillation is a
 natural assumption\cite{B.Hess,J.J.Tyson}.\\

We have carried out several simulations by taking a variety of
reaction networks that produce oscillatory dynamics.  In many of such examples, we have found the
differentiation process to be discussed. Here
we focus on such case, mostly using one typical example, by fixing a given network.
The oscillation of chemical
concentrations at the first stage in this adopted example is shown 
 as type 0 in Fig.4.  
This oscillation of chemical concentrations is
observed for most initial conditions,
although for rare initial condition there is also a fixed point solution whose
basin volume is very small.  Note that the oscillation of chemicals,
and accordingly the expressions of genes, show on/off type
switching, as is true in realistic cell systems.

Now we discuss the behavior of cells with the increase of cells.
As cells reproduce, the chemical state starts to be differentiated,
in consistency with the ``isologous diversification".
First, the phase coherence of oscillations is lost in the intra-cellular dynamics with the
increase of the number of cells. Then,
the chemical state of cells differentiates into 2 groups.  Each group has
a different composition in metabolites and also in other enzymes.  In the example
shown in Fig.5,  the type 1 cell is differentiated from the type 0 cell. Here
the type 0 cell has a higher activity with a larger metabolite concentrations,
than the type 1.  In order for a cell to grow, metabolites, enzyme, 
and ribonucleic acids are necessary.  The growth speed of a cell
depends on the balance among the concentrations  of 
chemicals $c_i^{(j)}$, $a_i^{(j)}$, $e_i^{(j)}$, and $x_i^{(j)}$ in our model.  
Hence the growth speed of
a cell also differentiates, depending on the concentration of
metabolites.  Since the dynamic states of chemicals are stabilized 
by the cell-cell interaction, these states, as well as the number ratio 
between the two types of cells, are stable against fluctuations.

The differentiation itself has already been studied in the earlier models
\cite{KK94,KY97,KY99,FK98,FK98b,KF00}. With the introduction of the transcription from tRNA, we
can discuss the difference in the use of genetic codes here.
Depending on the different metabolic states,
use of ARS is also differentiated.  In the present example,
the type 0 cell uses $e^{(1)}$, $e^{(5)}$, $e^{(7)}$, and $e^{(9)}$, 
and the concentrations of other ARS are zero.  On the other hand, the type 1 cell 
uses $e^{(5)}$ and  $e^{(7)}$ (see Fig.5). 
Therefore, each cell type has a different phenotype-genotype mapping.
Accordingly, we have found that different coding for the translation is adopted depending on the physiological
state of the cell.

When the type 1 cells are isolated, (i.e., by removing the type 0 cells), their
state switches to another type with distinct chemical composition.  This
type is called type 2. The type 2 cell uses $e^{(7)}$ only in all ARS.
In other words, the use of $e^{(5)}$ that is common to type 0 and type 1 cells
is abandoned, when the type 0 cells do not coexist.
This suggests that the adopted coding system may change depending on
cell-cell interaction(see Fig.6 for schematic representation).

\section{Evolutionary Process leading to Different Genetic Codes}

Now we consider the evolutionary process of the genetic code,
by introducing the change of the matrices
$U$ and $V$, giving the translation from 
nucleic acids to amino acids.  At every division a small noise is
introduced to $U$ and $V$ as mutation to genes. This noise corresponds to the fluctuation to the mapping between genotype and phenotype, and our purpose is to see how the evolution of coding progresses in the presence
of isologous diversification. 
 To include the selection process, cells are removed randomly, so that the total
number of cells is kept within a certain limit.  Since cells continue to
divide competing for resources, 
the selection process works as to the division speed of a cell.
Here the limit is set to be 150 in this simulation. 

First, we study the case when the phenotypes are not
differentiated in our model.  We choose a reaction network $W$
 so that the chemical dynamics fall onto a fixed point. Then, no
differentiation in cell types is observed.
In this case, even if the mutation to $U$ and $V$ is added, no
important change is observed.  
The values of matrix elements and chemical concentrations are
distributed with the variance given from the mutation rate, but
no differentiation to different groups of chemical states and
matrix elements (genotype parameters) is observed.  All the cells
keep adopting the same translation code from nucleic acid to amino acid 
and this coding does not change in time.

Since we are interested in the evolution of the code,
we do not adopt such network without differentiation.
In \S 2, existence of distinct physiological states
by isologous diversification is a necessary condition to have a distinct 
group in the genotype.
Hence, we adopt the network so that the chemical states of a cell
are differentiated to allow for different uses of ARS to synthesize
tRNA-amino acid complexes.  Accordingly, we choose the
matrix $W$ adopted in the previous section (for Fig.2).

Of course, the evolutionary process depends on the mutation rate, which is
given by the amplitude of noise added to the matrix elements $\varepsilon$.
If $\varepsilon$ is larger than $10^{-2}$, 
differentiation produced initially is destroyed, and the types are not preserved by cell divisions.
With such high mutation rate, the distribution of matrix elements by
cells is broader, and both the genotypes and phenotypes are
distributed without forming any distinct types. Then, the initial loose coupling
between genotype and phenotype remains.
No trend in the evolution of codes is observed.

When the mutation rate is lower, the genotypes, i.e., the
matrix elements also start to differentiate.
Each group with different compositions of metabolites starts to take
different matrix element values.
An example of the time course of some matrix elements is
shown in Fig.8 (a). 
Two separated groups are formed according to the differentiated
chemical states of metabolites given in Fig.7.  With the mutation and
selection process, the genotype is also differentiated 
following the phenotypic differentiation. This differentiation, originally brought about by the
interaction among cells, is embedded gradually in genotypic functions. 

Not all the elements of $U$ and $V$, but only some of them 
split.  In fact, metabolites or enzymes having
higher concentrations are often responsible for the differentiation. To estimate the splitting speed in the genotype space, 
we have plotted the distance of the values of an element
of $U$ and $V$ between the two types.  To be specific,
we have measured the following
distance between the averaged values of a given matrix element
of each type, i.e.,
\begin{eqnarray}
d^{(j,k)}=\mid 1/N_0\sum_{i\in type 0}^{N_0} S_i^{(j,k)}-1/N_1\sum_{i\in type 1}^{N_1} S_i^{(j,k)}\mid \nonumber
\end{eqnarray}
where S represents either V or U, and $N_0$ and $N_1$ are the number of type 0 and
1 cells  respectively.
As shown in Fig.8 (b), the separation progresses 
linearly in time, although  the mutational process is random.
In this sense, this separation process is rather fast and
deterministic in nature, once the phenotype is differentiated,
as is expected from the theory of \S 2.
Furthermore, the slope in the figure is different by chemicals,  although the
same mutation rate is adopted for all elements.  For some of other matrix elements, no separation occurs.

With this mutation process,  the difference in
chemical states is also amplified as shown in Fig.8.
With this evolutionary process, the differentiation starts to be more rigid.
In Fig.7, we have plotted the return map of the chemical states.
Now, the frequency of the differentiation event from type 0 to type 1
is decreased in time.  Each type keeps recursive production.

Next, we examine this separation process
by ``transplant" experiment, to see if each group of cells exists on its own.
At the initial stage of evolution, when 
type 0 cells are extracted, some of them spontaneously differentiate to
type 1 cells.  Type 0 cells cannot exist by themselves. 
With the evolution to change the genes, the rate of differentiation to
type 1 cells from isolated type 0 cells is
reduced.  Later at the evolution ($\sim$ 600 generation), 
transplanted type 0 cells no more differentiate 
to the type 1, and the type 0 cell stably exists on its own. 

On the other hand, as already mentioned, the type 1 cells, when transplanted, 
are transformed to the type 2 cells, where different ARS are used 
in the translation process (see Fig.9).  This characteristic feature
does not change through the evolution.

With this type of fixation process, the difference in the correspondence
between nucleic acid and protein (enzyme) is fixed.
For matrix $U$, one of the elements $U^{(i,j)}$ for given $j$
is larger through the evolution.   As shown in Fig.10, 
$U^{(7,0)}$ increases for the type 1 cell, with the decrease of $U^{(7,5)}$,
implying that the correspondence between $x^{(0)}$ and the ARS $e^{(5)}$ is stronger.
In other words, the loose correspondence
between the nucleic acid and amino acid is reduced in time, and a tight relationship between them is established.

Due to the evolution of matrix elements, the difference
in the correspondence between the type 0 and 1 cells gets
amplified.  Hence, the difference in the correspondence,
initially brought about as distinct metabolic states, is now
fixed into genes, and each type of cell, even after isolation,
keeps a different use of ARS for the translation of
the genetic information.

\section{Summary and Discussion}

In the present paper, we have studied how different correspondences between
nucleic acids and enzymes are formed and maintained through the
evolution.  To discuss this problem we have adopted a model
of a cell consisting of\\
(a) intracellular metabolic network\\
(b) ambiguity in translation system\\
(c) cell-cell interaction through the medium\\
(d) cell division\\
(e) mutation to the correspondence between nucleic acid and enzyme

According to our theory, the evolution of genetic code is
summarized as follows.

(1) First, due to the intracellular biochemical dynamics
with metabolites, enzymes, tRNA, and ARS, distinct
types of cells with distinct physiological states are formed
(for example, denoted by  type 0 and  type 1 cells).
Each cell type has different chemical composition and also uses different species of ARS 
for the protein synthesis.  Hence, each cell type adopts different correspondence between nucleic acids and enzymes. 

The differentiation at this stage is due to cell-cell interaction.
For example, a type 1 cell is differentiated from a type 0 cell, 
and can maintain itself only under the presence of type 0 cells. 
The difference in the correspondence, however, is
not fixed as yet, and by each cell division, each cell can take a different
metabolic state, and the correspondence is changeable.

(2) Next, by mutation to the catalytic ability of enzyme 
by each division of a cell,  each distinct cell type starts to be
fixed, and keeps its type after the division.
The difference in chemical states is now
fixed to parameters that characterize the catalytic ability.
Accordingly, each cell type with distinct
metabolic states is fixed also to the catalytic ability of enzymes
represented by genes.

(3) After the fixation of distinct types is completed both in phenotypes and genotypes, 
these types are maintained even if each type of cells is  isolated.  Each type
uses different ARS for the translation
between nucleic and amino acids, and this difference in the usage is 
amplified through the evolution.  At this stage, one can say that
different coding, originally introduced as distinct physiological states of cells
through cell-cell interaction, is established genetically.

The presented result here is rather general, as long as cellular states 
differentiate into a few types, as is generally observed in a model adopting 
the processes (a)-(e).
Although the network we have adopted is randomly chosen, it is expected that the
same evolution process of genetic code is observed as long as this
general setup with (a)-(e) is satisfied.
As mentioned in \S 2, this evolutionary process here is based on the 
standard Darwinian process without any Lamarckian mechanism, 
although the genetic fixation
occurs later from the phenotypic differentiation. 

\subsection{The origin of mitochondrial non-universal genetic code}

Our theory of the evolution of genetic code can shed new light on
the non-universal genetic code of mitochondria.  From recent studies in the
molecular biology, it is suggested that mitochondria had used almost the 
same code as 
universal one before ``endosymbiosis"\cite{L.Margulis}, and 
its genetic code was deviated after symbiosis\cite{S.Osawa}.  

According to our theory for the evolution of genetic code,
the coding system can depend on cell-cell interaction. A type 2 cell, that is 
formed by the isolation of a type 1 cell, has a different use of ARS than a cell in
coexistence with the type 1 cell.  With the interaction, the cells take
a different coding system.   Furthermore, this difference in the coding
is established through the evolution.
In this sense, it is a natural course of evolution that mitochondria,
which starts to live within a cell and has strong interaction with the host cell, 
will establish a different coding system through the evolution.

Although the evolution to switch to a different coding might look fatal to an organism,
a cell can survive via the loose coupling between the
genotype and phenotype.  The loose coupling produced by the cell-cell
interaction is essential to the evolution to non-universal genetic codes.

It should also be stressed that the genetic code is not necessarily solely determined
by the genetic system.  In a biochemically  plausible model, we have demonstrated
that the change in the physiological state of a cell can lead to difference in the genetic code.  
Based on our theory  we believe that this dependence on the physiological 
state is essential to the study of non-universal coding in  mitochondria 
and others. Furthermore, such possibility of the difference in coding may 
not be limited to the phenomena at the early stage of evolution.  It may be 
possible to pursue such possibility experimentally, by changing the nature of
interaction among cells or intracellular organs keeping genetic information.

\subsection{Relevance to artificial life}

Discussion of the mechanism involved in evolution often 
remains vague, since no one knows for sure what has occurred
in history, within limited fossil data.
Most important in our theory, on the other hand,
lies in experimental verifiability. 
As mentioned, isologous diversification has already been
observed in the differentiation of enzyme activity of {\it E. coli} with
identical genes\cite{Yomo,Kashi}.  We have already started an experiment
of the evolution of {\it E. coli} in
the laboratory\cite{Xu}, controlling the strength of the 
interaction through the population density.  
With this experiment we can check if the evolution on 
the genetic level is accelerated through
interaction-induced phenotypic diversification,
and can answer if the evolution theory of \S 2 really occurs in nature.
In this sense, our theory is testable in laboratory,
in contrast with many other speculations.
Change of genetic code through evolution can also be checked in laboratory.

In the same sense, our study is relevant to the field of artificial life (AL), 
since AL attempts to
understand some biological process such as  evolution, by constructing an 
artificial system in laboratory or in a computer from our side.  

A problem in most of the present AL studies lies in that
it is too much symbol-based.  They generally assume some rule, 
represented as manipulation over symbols in the beginning.
A model by such rules will eventually be written by a universal Turing machine.  Hence
it generally faces with the problem that the emergence may not be possible
in principle in such system, since the emergence originally
means a generation of a novel, higher level that is not originally
written in a rule.
The same drawback lies in the symbol-based study of evolution
(i.e., a study starting from the evolution of symbols corresponding to
genes), and indeed, the AL study on the evolution is often nothing but
a kind of complicated optimization problem.

According to our theory, first the phenotype is differentiated, 
given by continuous (analogue) dynamical system, which is later fixed to genes
that serve as a rule for dynamical systems.  Now, rules written by
symbols (genetic codes) are not necessarily the principal
cause of the evolution\cite{Complexity}.
~\\

We would like to thank C.Furusawa for stimulating discussions.
This work is supported by
Grant-in-Aids for Scientific Research from
the Ministry of Education, Science and Culture of Japan
(11CE2006;Komaba Complex Systems Life Project; and 11837004).
~\\

\clearpage

\begin{figure}
\begin{center}
\includegraphics[width=13cm,height=13cm]{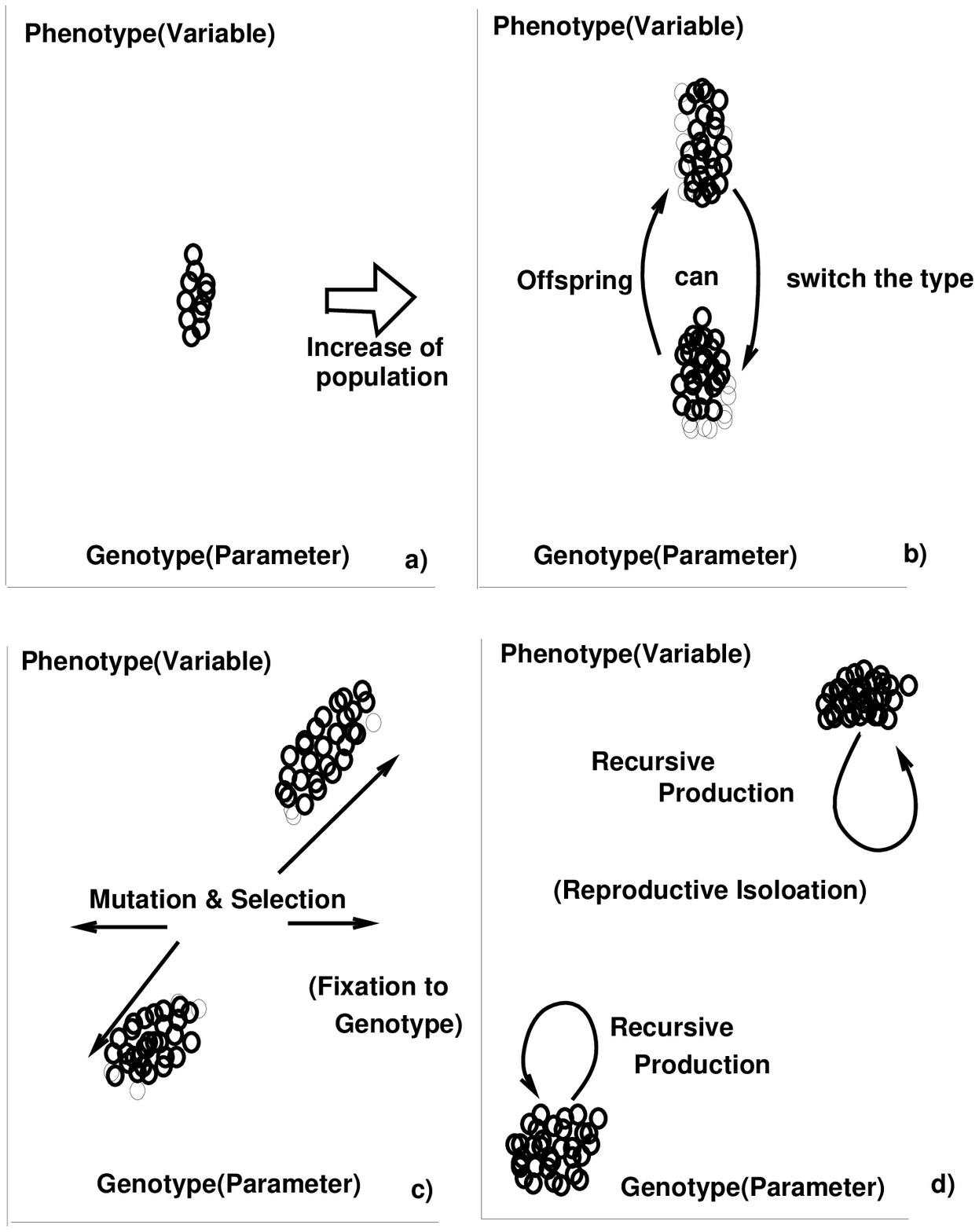}
\vskip 0.25cm
\caption{Schematic representation of speciation process, plotted as phenotype-genotype relationship.
(a) Initially, there is a group of organisms with distribution
centered around a given phenotype and genotype.
(b) Then, with the increase of population, phenotype is differentiated 
into discrete types. (c) Then according to the difference of
phenotype, genotype is also differentiated.  (d) Finally, the two
groups differentiate both in genotypes and phenotypes, and form
distinct species.  Indeed, these two groups are separated also
by sexual recombination, since the hybrid offspring cannot produce its progeny.}
\end{center}
\end{figure}

\clearpage

\begin{figure}
\begin{center}
\includegraphics[width=15cm,height=8cm]{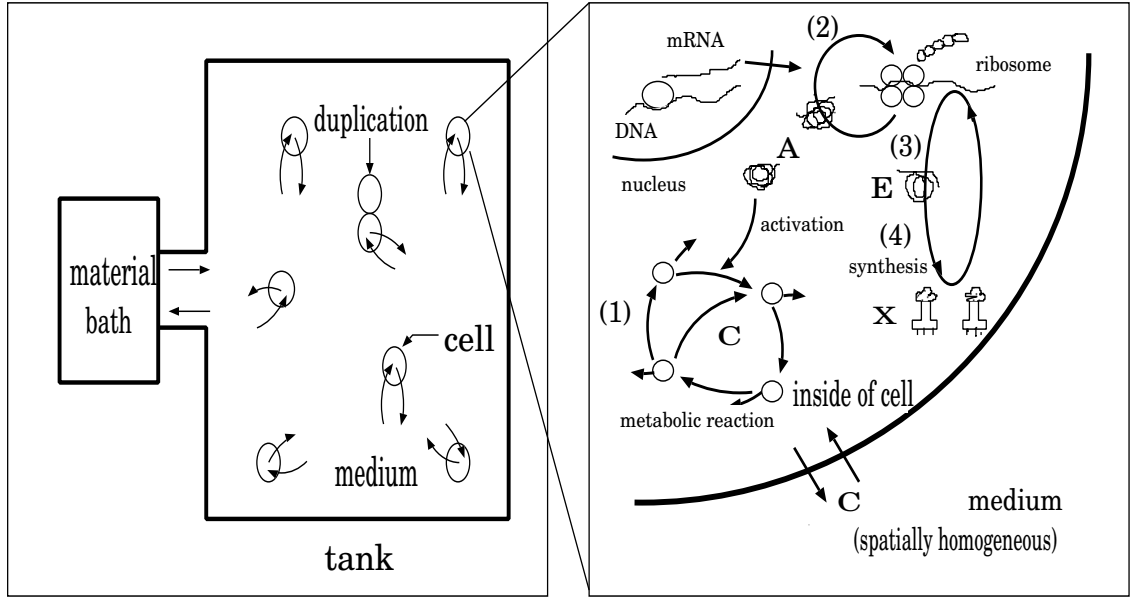}
\caption{Schematic representation of our model} 
\end{center}
\end{figure}

\begin{figure}
\begin{center}
\includegraphics[width=12cm,height=7cm]{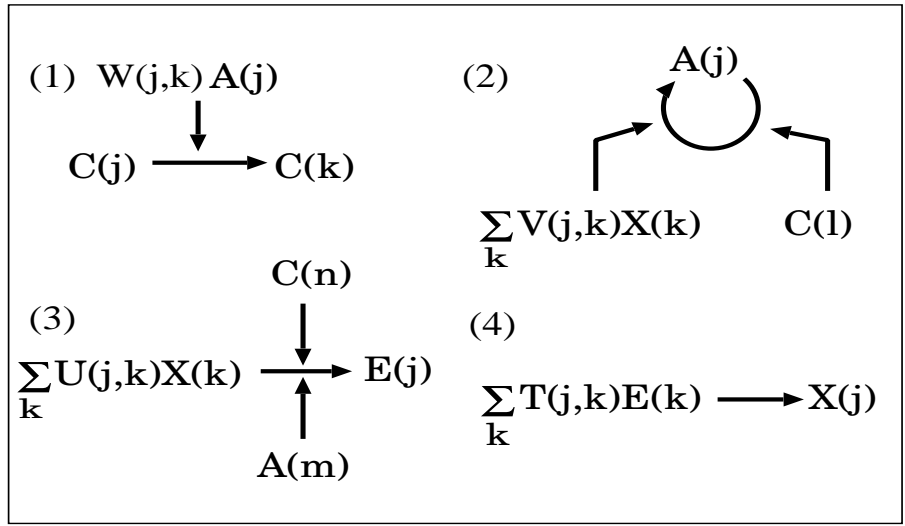}
\caption{Schematic representation of intra-cellular reaction network} 
\end{center}
\end{figure}

\clearpage

\begin{figure}
\begin{center}
\includegraphics[width=10cm,height=15cm]{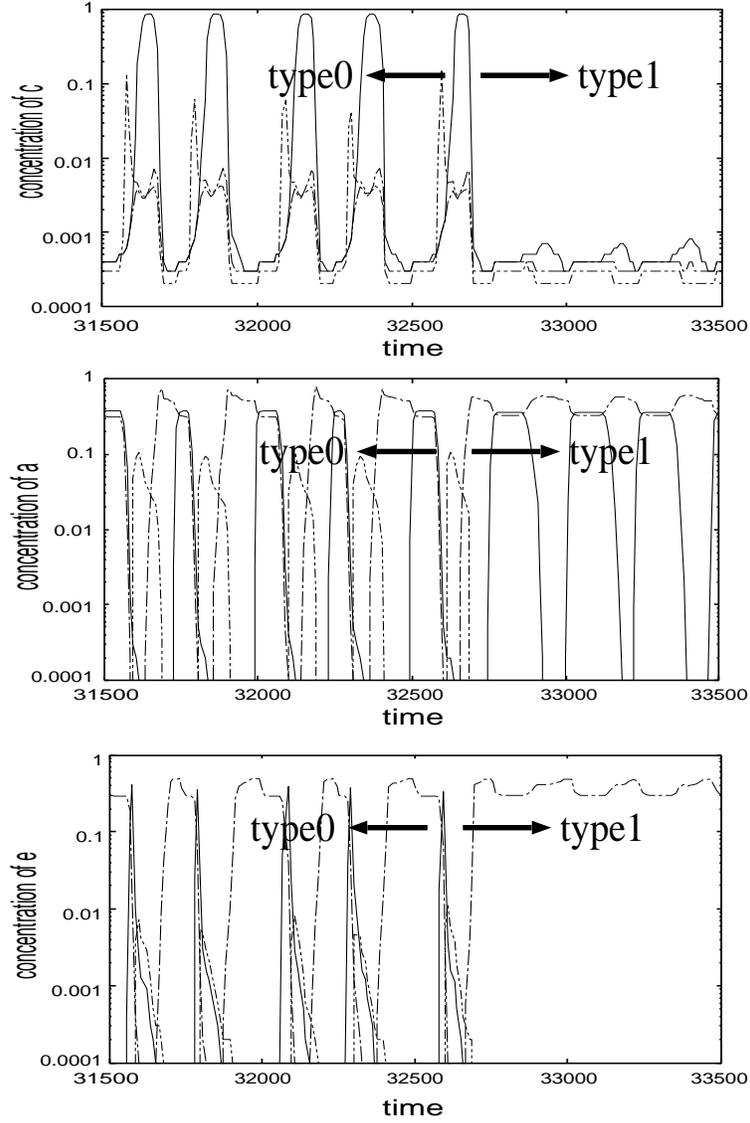}
\caption{Oscillations of chemicals. The time series of some $c_i^{(j)}$, $a_i^{(j)}$, $e_i^{(j)}$ are plotted by semi-log scale. The parameters are set at
$D_1=3.0$, $D_2=0.050$, $D_3=100.0$, $D_4=100.0$, $D_5=1.0$, $D_6=0.050$, $Vol=100.0$, $\overline{C^{(j)}}=0.010$ (for all j) in all the simulations
shown in the present paper.} 
\end{center}
\end{figure}

\clearpage

\begin{figure}
\begin{center}
\includegraphics[width=9cm,height=5cm]{fig5.eps}
\includegraphics[width=9cm,height=5cm]{fig6.eps}
\caption{The return map of the average concentration of $a^{(12)}$ 
(a), 
and the plot of the average concentration of ($e^{(1)}$,$e^{(7)}$) of
each cell (b),
plotted at every division event. In the return map, the chemical average of
a mother cell as abscissa and that of its daughter cell as ordinate are plotted.
As shown, the type 1 cell keeps its type after division, while the type
0 cell either proliferates or differentiates to type 1.} 
\vskip 0.5cm
\includegraphics[width=9cm,height=5cm]{fig7.eps}
\caption{Schematic representation  of the differentiation to the types
observed in our model} 
\end{center}
\end{figure}

\clearpage

\begin{figure}
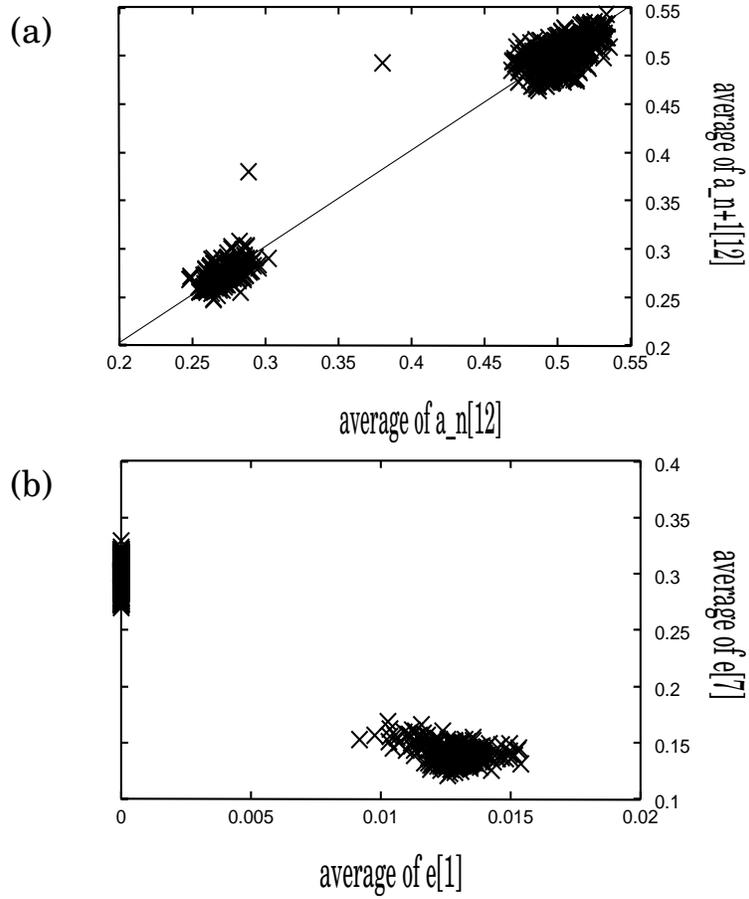

\begin{center}
\includegraphics[width=10cm,height=6cm]{fig8.eps}
\includegraphics[width=10cm,height=6cm]{fig9.eps}
\caption{(a) is the
return map of the average concentration of $a^{(12)}$, 
plotted in the same way as Fig.5, while (b) shows the
plot of the average of ($e^{(1)}$ ,$e^{(9)}$), plotted at every division event.
Each cell keeps recursive production.} 
\end{center}
\end{figure}

\begin{figure}
\begin{center}
\includegraphics[width=10cm,height=6cm]{fig10.eps}
\includegraphics[width=10cm,height=6cm]{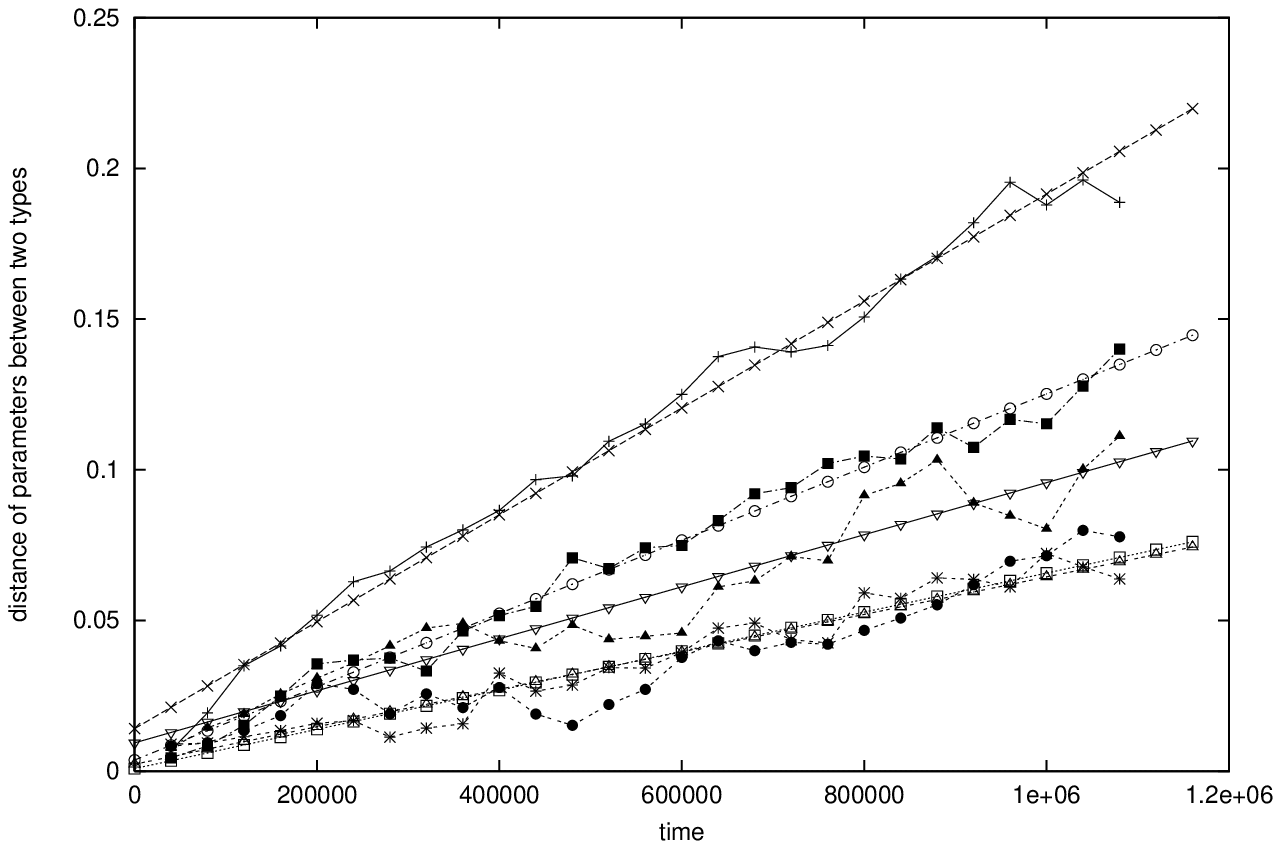}
\caption{The temporal change of $V^{(12,2)}$,
and $V^{(12,5)}$, namely, the activity of $a^{(12)}$ for the composition $x^{(2)}$, and $x^{(5)}$ (a). The parameter values are plotted at each division event. 
The temporal change of the distance between 
the averages for the matrix elements of each type is shown in (b).}
\end{center}
\end{figure}

\clearpage

\begin{figure}
\begin{center}
\includegraphics[width=12cm,height=7cm]{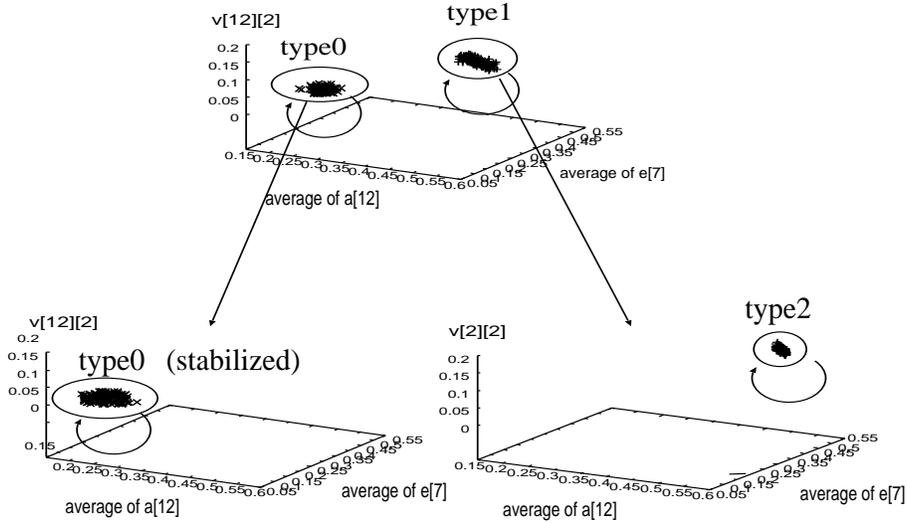}
\caption{Genotype-phenotype relation after the transplant experiment.
The set of values  $(a^{(12)},e^{(7)},V^{(12,2)})$ is plotted at every division event.}
\end{center}
\end{figure}

\begin{figure}
\begin{center}
\includegraphics[width=10cm,height=7cm]{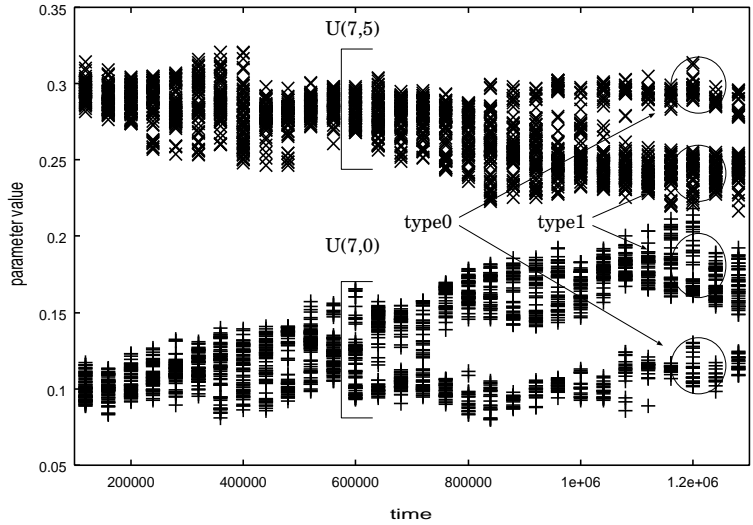}
\caption{The change of the matrix $U$ through the evolution.
The parameter values  $U^{(7,0)}$ and $U^{(7,5)}$ are
plotted at each division event.  First, each type
starts to take different  $U^{(7,0)}$ values and later
$U^{(7,5)}$ values.  For example, the type 1 cell starts to use more $x^{(0)}$
and $e^{(7)}$.} 
\end{center}
\end{figure}

\end{document}